\documentclass{jnmp}
\usepackage{amsmath}
\usepackage{bm}
\usepackage{graphicx}

\JNMPnumberwithin{equation}{section}

\theoremstyle{definition}

\newcommand{\eqdef}{\stackrel{\rm def}{=}}
\newcommand{\sfrac}[2]{{\textstyle \frac{#1}{#2}}}

\begin{document}

%
\renewcommand{\evenhead}{S Odake \& R Sasaki}
\renewcommand{\oddhead}{Shape Invariant Potentials in
``Discrete Quantum Mechanics"}

%
\thispagestyle{empty}



\vspace*{-15mm}
\newfont{\elevenmib}{cmmib10 scaled\magstep1}
\begin{flushleft}
  \elevenmib Yukawa\, Institute\, Kyoto\\
\end{flushleft}\vspace{-1.3cm}
\begin{flushright}\normalsize  \sf
  DPSU-04-3\\
  YITP-04-55\\
  September 2004\\
  {\tt hep-th/0410102} 
\end{flushright}

\Name{Shape Invariant Potentials in ``Discrete Quantum Mechanics"
\footnote{
Contribution to a special issue of Journal of Nonlinear Mathematical Physics
in honour of Francesco Calogero on the occasion of his seventieth birthday.
}
}

\label{firstpage}

\Author{Satoru  ODAKE~$^\dag$ and Ryu  SASAKI~$^\ddag$}

\Address{$^\dag$ Department of Physics, Shinshu University,
     Matsumoto 390-8621, Japan \\
~~E-mail: odake@azusa.shinshu-u.ac.jp\\[10pt]
$^\ddag$ Yukawa Institute for Theoretical Physics,
     Kyoto University, Kyoto 606-8502, Japan \\
~~E-mail: ryu@yukawa.kyoto-u.ac.jp}


\begin{abstract}
\noindent
Shape invariance is an important ingredient  of many exactly solvable
quantum mechanics. Several examples of shape invariant ``discrete quantum
mechanical systems" are introduced and discussed in some detail. They arise
in the problem of describing the equilibrium positions of
Ruijsenaars-Schneider type systems, which are ``discrete" counterparts of
Calogero and Sutherland systems, the celebrated exactly solvable
multi-particle dynamics. Deformed Hermite and Laguerre polynomials are the
typical examples of the eigenfunctions of the above shape invariant
discrete quantum mechanical systems.
\end{abstract}

%
\section{Introduction}
\label{intro}
Many exactly solvable quantum mechanical systems, for example,
the harmonic oscillator without/with a centrifugal potential,
the coulomb problem, the trigonometric and hyperbolic P\"oschl-Teller
potential, the symmetric top {\em etc.\/}
are {\em shape invariant\/}\cite{genden}.
Here we will show that this concept is also useful in a wider context of
``discrete quantum mechanics", in which the momentum operator $p=-i\hbar
d/dx$ appears as an exponential (hyperbolic) function instead of a
polynomial in ordinary quantum mechanics.
For demonstration, we present several explicit examples of shape invariant
discrete quantum mechanics in some detail.
The corresponding eigenfunctions are orthogonal polynomials belonging to the
family of Askey-scheme of hypergeometric orthogonal polynomials
\cite{And-Ask-Roy,koeswart}.
The examples, roughly speaking, are related to one and two parameter
deformation of the Hermite polynomials and two and three parameter
deformation  of the Laguerre polynomials.
They could be interpreted as describing deformed harmonic oscillator
with (Laguerre) and without (Hermite) the centrifugal potential.
They also arise in many contexts of theoretical physics 
\cite{atasus,benderkoorn}.

These examples arise within the context of multi-particle exactly solvable
quantum mechanical systems, in particular, the Calogero and Sutherland
systems \cite{Cal-Sut} and their deformation called the
Ruijsenaars-Schneider-van Diejen systems \cite{RSvD}.
Their integrability is based on the root systems and the associated Weyl
(Coxeter) group symmetry.
More than twenty years ago, Calogero \cite{calmat}
found out that the equilibrium position of the A-type Calogero system
was described by the zeros of the Hermite polynomial.
The same fact was discussed by Stieltjes in a slightly different context
more than a century ago \cite{sti}.
The equilibria of the B, BC and D type Calogero systems are described by the
zeros of the Laguerre polynomials.
The equilibria of the Sutherland (trigonometric potential) systems based on
the classical root systems are described by the zeros of the Chebyshev and
Jacobi polynomials. The interesting dynamics associated with  the equilibria
of exactly solvable multi-particle systems and the polynomials describing  the
equilibria of Calogero-Sutherland (C-S) systems based on the exceptional root
systems are presented in \cite{cs}.
Ruijsenaars-Schneider (R-S) and van Diejen introduced integrable deformation of
the C-S systems, with several additional parameters
\cite{RSvD}. The equilibrium positions of the R-S and van
Diejen systems are discussed by Ragnisco-Sasaki and Odake-Sasaki
\cite{rsos2}. As expected certain multi-parameter deformation of the Hermite
and Laguerre polynomials describe the equilibrium positions of the rational
R-S or the van Diejen systems.

The single particle dynamics related to these polynomials is the main
subject of this paper. They are certain
{\em deformation\/} of the harmonic oscillator without/with the centrifugal
potential belonging to the ``discrete quantum mechanics".
We show that they are exactly solvable thanks to the inherited shape
invariance.
To the best of our knowledge, they are the first examples of shape invariance
discussed in the framework of discrete quantum mechanics.
Of course there are plenty of works applying the {\em factorisation\/} method
\cite{infhul,crum,susyqm} to various {\em difference\/} equations 
\cite{discrete,spivinzhed}.
However, all of these difference equations contain finite shifts in the {\em real\/}
direction, which result in polynomials of a {\em discrete
variable\/}, for example the Charlier, the Meixner and the Hahn
polynomials \cite{And-Ask-Roy}. Apparently they look  very
different from those occurring in quantum mechanics.

This paper is organised as follows.
In section two, the simplest example of a shape invariant discrete quantum
mechanics, related to a deformed harmonic oscillator is introduced and
discussed in some detail. The eigenfunctions are deformed Hermite
polynomials, or a special case of the Meixner-Pollaczek polynomials
\cite{koeswart}.
The course of the arguments is essentially the same as in ordinary quantum
mechanics or the Sturm-Liouville problem, as most clearly formulated in the
seminal work of Crum \cite{crum}.
It is also known as the factorisation method \cite{infhul}, or the
supersymmetric quantum mechanics \cite{susyqm}.
Here we follow  the notion and notation of Crum's paper.
In section three we present three explicit examples of shape invariant
discrete quantum mechanics, in which a two parameter deformation of the Hermite
polynomials, a two and three parameter deformation of Laguerre polynomials
are the eigenfunctions. They are a special case of the continuous Hahn
polynomial,
the continuous dual Hahn polynomial and the Wilson polynomial \cite{koeswart}.
Shape invariance is demonstrated elementarily.
The meaning of the factorisation at the level of the Hamiltonian and at the
level of the polynomial solutions are elaborated in  detail.
Although these polynomials are perfectly well understood in their own right, we
do believe {\em dynamical\/} interpretation in terms of the
factorised Hamiltonian, the analogues of the creation and the annihilation
operators, {\em etc.\/} would shed new light on them.
Section four is devoted to the clarification of the background of the present
research: how these polynomials came to our notice through  the polynomials
describing the equilibrium points of the van Diejen systems based on the
classical root systems.
The final section is for comments, a summary and an outlook.

\section{Deformed Hermite (Meixner-Pollaczek) Polynomials}
\label{defher}

The simplest and best known example of {\em exactly solvable\/}
and {\em shape invariant\/}
 quantum mechanics is the harmonic oscillator
\begin{equation}
  H=-\frac{1}{2}{\partial^2_x}+\frac{1}{2}(x^2-1) 
=\frac{1}{2}\left(-i\partial_x+ix\right)\left(-i\partial_x-ix\right)
  ,\qquad \partial_x\eqdef \frac{d}{dx}.
  \label{osciham}
\end{equation}
The Hamiltonian  is factorised and the eigenfunctions are the
Hermite polynomials,
$H_n(x)$:
\begin{equation}
  H\phi_n=\mathcal{E}_n\phi_n,\quad
  \mathcal{E}_n=n,\quad
  \phi_n(x)\propto H_n(x)\,e^{-x^2/2}, \quad n=0,1,\ldots, .
\end{equation}

One naively expects that a certain {\em deformation\/}
of the Hermite polynomials would constitute the
eigenfunctions of a simplest {\em shape invariant\/}
``discrete quantum mechanics".
This is exactly the case and will be explained in some detail in this section.

Let us start with the following Hamiltonian,
\begin{gather}
  H\eqdef\sfrac12\sqrt{V(x)}\,e^{-i\partial_x}\sqrt{V^*(x)}
  +\sfrac12\sqrt{V^*(x)}\,e^{i\partial_x}\sqrt{V(x)}
  -\sfrac12(V(x)+V^*(x)),
  \label{H}\\[4pt]
 V(x)=\lambda+i x,\qquad V^*(x)=\lambda-i x\,;\qquad \lambda\in\mathbb{R}_+.
\end{gather}
In the discrete quantum mechanics the momentum operator
$p=-i\hbar {\partial_x}$ (with $\hbar=1$)
appears as exponentiated instead of powers
in ordinary quantum mechanics.
Thus they cause a finite shift of the wavefunction in the {\em imaginary\/}
direction. For example,
\begin{equation*}
  e^{\pm i\partial_x}\phi(x)=\phi(x\pm i).
\end{equation*}
Throughout this paper we adopt
the following convention of a complex conjugate function:
for an arbitrary function $f(x)=\sum_na_nx^n$, $a_n\in\mathbb{C}$ we define
$f^*(x)=\sum_na_n^*x^n$. Here $c^*$ is the complex conjugation of a number 
$c\in\mathbb{C}$.
Note that $f^*(x)$ is not the complex conjugation of
$f(x)$, $(f(x))^*=f^*(x^*)$.
This is particularly important when a function is shifted in the imaginary
direction.

\paragraph{Factorised Hamiltonian}
It is easy to see that the above Hamiltonian (\ref{H}) is {\em factorised\/}:
\begin{gather}
  H=A^{\dagger}A,
  \label{Hfac}\\
  A=A(x;{\lambda})
  \eqdef\frac{1}{\sqrt{2}}\Bigl(e^{-\frac{i}{2}\partial_x}\sqrt{V^*(x)}
  -e^{\frac{i}{2}\partial_x}\sqrt{V(x)}\Bigr),\\
  A^{\dagger}=A(x;{\lambda})^{\dagger}
  \eqdef\frac{1}{\sqrt{2}}\Bigl(\sqrt{V(x)}\,e^{-\frac{i}{2}\partial_x}
  -\sqrt{V^*(x)}\,e^{\frac{i}{2}\partial_x}\Bigr).
\end{gather}
Here $\dagger$ denotes the ordinary hermitian conjugation
with respect to the ordinary $L^2$
inner product:  $(f,g)=\int_{-\infty}^\infty(f(x))^*g(x)dx$.
Obviously the Hamiltonian (\ref{H}) is hermitian
(self-conjugate) and positive semi-definite.

The ground state $\phi_0$ is annihilated by $A$:
\begin{equation}
  A\phi_0(x)=0,\quad \phi_0(x)\propto
  \sqrt{\Gamma(\lambda+ix)\Gamma(\lambda-ix)}\quad
  \Longrightarrow H\phi_0(x)=0,\quad \mathcal{E}_0=0.
  \label{phi0form}
\end{equation}
The above equation reads
\begin{equation}
  \sqrt{V^*(x-\sfrac{i}{2})}\,\phi_0(x-\sfrac{i}{2})
  =\sqrt{V(x+\sfrac{i}{2})}\,\phi_0(x+\sfrac{i}{2}).
\end{equation}

The other eigenfunctions of the Hamiltonian (\ref{H}) can be obtained
in the form
\begin{equation}
  H\phi_n=\mathcal{E}_n\phi_n,\quad
  \phi_n(x)\propto P_n(x) \phi_0(x) \quad \Longrightarrow
  \tilde{H}P_n=\mathcal{E}_n P_n,
  \label{Htileq}
\end{equation}
in which $\tilde{H}$ is a similarity transformed
Hamiltonian in terms of the ground state wavefunction $\phi_0$:
\begin{gather}
  \tilde{H}\eqdef\phi_0^{-1}\circ H\circ\phi_0\nonumber\\
  \quad \ =\sfrac12V(x)e^{-i\partial_x}+\sfrac12V^*(x)e^{i\partial_x}
  -\sfrac12(V(x)+V^*(x)).
  \label{tilH}
\end{gather}
The eigenvalue equation (\ref{Htileq}) is now the
following {\em difference\/} equation
\begin{equation}
  (\lambda+ix)P_n(x-i)+(\lambda-ix)P_n(x+i)=
  2(\lambda+\mathcal{E}_n)P_n(x),
\end{equation}
which obviously has a polynomial solution,
$P_n(x)=a_n x^n+$ lower degree terms, of {\em definite parity\/},
$P_n(-x)=(-1)^nP_n(x)$. By comparing the coefficient of the highest degree
$x^n$ term, we find easily
\begin{equation}
  \mathcal{E}_n=n, \quad n=1,2,\ldots, .
\end{equation}
The orthogonal polynomials $\{P_n(x), n\ge0\}$
with respect to the weight function
\begin{equation}
  \omega(x;\lambda)=
  \Gamma(\lambda+ix)\Gamma(\lambda-ix)\propto \phi_0(x)^2
\end{equation}
satisfy the {\em three term recurrence\/}, which reads
\begin{equation}
  p_{n+1}(x)=x\,p_n(x)-{n(n+2\lambda-1)/4}\,p_{n-1}(x),
  \quad n\ge0,\quad p_0=1,\quad p_{-1}=0,
\end{equation}
for the {\em monic\/} polynomial, $p_n(x)=x^n+$ lower degree terms.
With proper normalisation
\begin{equation} 
P_n(x)=\frac{2^n}{n!}p_n(x)=P^{(\lambda)}_n(x;\frac{\pi}{2}),
\end{equation}
it is a special case of the Meixner-Pollaczek polynomial.
Here we follow the notation of Koekoek and Swarttouw \cite{koeswart}.
Since the second argument $\frac{\pi}{2}$ remains fixed
throughout our discussion, we will omit it:
\begin{equation*}
  P^{(\lambda)}_n(x;\frac{\pi}{2})\eqdef P^{(\lambda)}_n(x).
\end{equation*}
They are polynomials in both $x$ and $\lambda$ with
{\em real\/} and {\em rational\/} coefficients.
The Meixner-Pollaczek polynomial also appears in a ``relativistic
oscillator" model
\cite{atasus}.

Corresponding to the factorisation of $H$ (\ref{Hfac}),
$\tilde{H}$ is also factorised:
\begin{gather}
  \tilde{H}=BC,\\
  C=C(x)=\sfrac{i}{2}(e^{-\frac{i}{2}\partial_x}
  -e^{\frac{i}{2}\partial_x}),\\
  B=B(x;{\lambda})=-i(V(x)e^{-\frac{i}{2}\partial_x}
  -V^*(x)e^{\frac{i}{2}\partial_x})=
  -i\left((\lambda+ix)e^{-\frac{i}{2}\partial_x}
  -(\lambda-ix)e^{\frac{i}{2}\partial_x}\right).
\end{gather}
They factorise the eigenvalue equation
\begin{equation}
  BC\,P_n^{(\lambda)}(x)=n\,P_n^{(\lambda)}(x)\
  \Longleftarrow \
  C\,P_n^{(\lambda)}(x)=P_{n-1}^{(\lambda+\frac{1}{2})}(x),\quad
  B\,P_{n-1}^{(\lambda+\frac{1}{2})}(x)=n\,P_n^{(\lambda)}(x),
\end{equation}
giving  rise to the {\em forward\/} and {\em backward\/} shift relations.
They read explicitly:
\begin{gather}
  C:\quad
  P_n^{(\lambda)}(x+\sfrac{i}{2})-P_n^{(\lambda)}(x-\sfrac{i}{2})
  =2i\,P_{n-1}^{(\lambda+\frac{1}{2})}(x),\\
  B:\quad
  i\left((\lambda-ix)P_{n-1}^{(\lambda+\frac{1}{2})}(x+\sfrac{i}{2})
  -(\lambda+ix)P_{n-1}^{(\lambda+\sfrac{1}{2})}(x-\sfrac{i}{2})
  \right)=n\,P_n^{(\lambda)}(x).
\end{gather}

Let us define a new set of wavefunctions
\begin{equation}
  \phi_{1,n}\eqdef A\phi_n,\quad n=1,2,\ldots, .
\end{equation}
As a consequence of the factorisation,
they form eigenfunctions of a new Hamiltonian
\begin{equation}
  H_1=AA^\dagger
\end{equation}
with the same eigenvalues $\{\mathcal{E}_n\}$:
\begin{equation}
  H_1\phi_{1,n}=AA^\dagger A\phi_n=A\mathcal{E}_n\phi_n
  =\mathcal{E}_n\phi_{1,n},\quad n=1,2,\ldots, .
\end{equation}

\paragraph{Shape Invariance}
It is straightforward to evaluate the reversed order product $AA^\dagger$:
\begin{gather}
  H_1=\sfrac12\sqrt{V_1(x)}\,e^{-i\partial_x}\sqrt{V_1^*(x)}
  +\sfrac12\sqrt{V_1^*(x)}\,e^{i\partial_x}\sqrt{V_1(x)}
  -\sfrac12(V_1(x)+V_1^*(x))+\mathcal{E}_1,
  \label{H1def}\\[4pt]
  V_1(x)=\lambda+\sfrac{1}{2}+i x,\qquad
  V_1^*(x)=\lambda+\sfrac{1}{2}-i x,\qquad \mathcal{E}_1=1.
\end{gather}
It has the {\em same form\/} as $H$ with a constant term
$\mathcal{E}_1=1$ added and the parameter $\lambda$ shifted by
$\frac{1}{2}$. Thus it is again factorised:
\begin{gather}
  H_1=AA^{\dagger}=A_1^{\dagger}A_1+{\cal E}_1,
  \label{H1fac}\\
  A_1=A_1(x;{\lambda})
  \eqdef\frac{1}{\sqrt{2}}\Bigl(e^{-\frac{i}{2}\partial_x}\sqrt{V_1^*(x)}
  -e^{\frac{i}{2}\partial_x}\sqrt{V_1(x)}\Bigr)=A(x;\lambda+\sfrac{1}{2}),\\
  A_1^{\dagger}=A_1(x;{\lambda})^{\dagger}
  \eqdef\frac{1}{\sqrt{2}}\Bigl(\sqrt{V_1(x)}\,e^{-\frac{i}{2}\partial_x}
  -\sqrt{V_1^*(x)}\,e^{\frac{i}{2}\partial_x}\Bigr)
  =A^{\dagger}(x;\lambda+\sfrac{1}{2}).
\end{gather}
The ground state wavefunction is $\phi_{1,1}$ annihilated by $A_1$:
\begin{gather}
  A_1\phi_{1,1}(x)=0,\quad \phi_{1,1}(x)\propto
  \sqrt{\Gamma(\lambda+\sfrac{1}{2}+ix)\Gamma(\lambda+\sfrac{1}{2}-ix)}\\
  \qquad\qquad\  \Longrightarrow H_1\phi_{1,1}(x)=\mathcal{E}_1\phi_{1,1}(x),
  \quad \mathcal{E}_1=1.
\end{gather}

Clearly this process can be repeated as many
times as the number of discrete levels of the original Hamiltonian $H$
(\ref{H}):
\begin{equation}
  H=H_0\to H_1\to H_2 \cdots \to H_s \to \cdots .
\end{equation}
All these Hamiltonians share the same spectra $\{\mathcal{E}_n\}$,
$n=0,1,\ldots$.
The eigenfunction of the $s$-th Hamiltonian $H_s$ are given by
\begin{equation}
  H_s\phi_{s,n}=\mathcal{E}_n\phi_{s,n},\quad
  \phi_{s,n}\eqdef  A_{s-1}A_{s-2}\cdots A_1 A\,\phi_n,\quad 0\leq s\le n,
  \label{phinsform}
\end{equation}
in which
\begin{gather}
  H_s=A_{s-1}A_{s-1}^{\dagger}+{\cal E}_{s-1}=A_s^{\dagger}A_s+{\cal E}_s,
  \label{Hs}\\
  A_s=A_s(x;{\lambda})
  \eqdef\frac{1}{\sqrt{2}}\Bigl(e^{-\frac{i}{2}\partial_x}\sqrt{V_s^*(x)}
  -e^{\frac{i}{2}\partial_x}\sqrt{V_s(x)}\Bigr)=A(x;\lambda+\sfrac{s}{2}),\\
  A_s^{\dagger}=A_s(x;{\lambda})^{\dagger}
  \eqdef\frac{1}{\sqrt{2}}\Bigl(\sqrt{V_s(x)}\,e^{-\frac{i}{2}\partial_x}
  -\sqrt{V_s^*(x)}\,e^{\frac{i}{2}\partial_x}\Bigr)
  =A(x;\lambda+\sfrac{s}{2})^{\dagger},\\
  V_s(x)=\lambda+\sfrac{s}{2}+i x,\qquad
  V_s^*(x)=\lambda+\sfrac{s}{2}-i x,\qquad \mathcal{E}_s=s,\\
  A_s\phi_{s,s}=0\quad \phi_{s,s}(x)\propto
  \sqrt{\Gamma(\lambda+\sfrac{s}{2}+ix)\Gamma(\lambda+\sfrac{s}{2}-ix)}\\
  \qquad\qquad\  \Longrightarrow H_s\phi_{s,s}(x)=\mathcal{E}_s\phi_{s,s}(x),
  \quad \mathcal{E}_s=s.
\end{gather}

By multiplying $A_s^\dagger$ to $\phi_{s+1,n}\eqdef A_s\phi_{s,n}$
we obtain
\begin{equation}
  A_s^\dagger \phi_{s+1,n}=A_s^\dagger A_s \phi_{s,n}
  =(H_s-\mathcal{E}_s)\phi_{s,n}=(\mathcal{E}_n-\mathcal{E}_s)\phi_{s,n}.
  \label{adagrel}
\end{equation}
Since the ground state $\phi_{n,n}$ of the $n$-th Hamiltonian
$H_n$ is known explicitly,
we can express the $n$-th eigenfunction $\phi_n\equiv \phi_{0,n}$ of the
original Hamiltonian $H$ (\ref{H}) in terms of $A^\dagger$ and $\phi_{n,n}$
by repeated use of the above formula (\ref{adagrel}):
\begin{equation}
  \phi_n(x;\lambda)
  =\frac{A_0^{\dagger}}{{\cal E}_n-{\cal E}_0}\,
  \frac{A_1^{\dagger}}{{\cal E}_n-{\cal E}_1}\cdots
  \frac{A_{n-1}^{\dagger}}{{\cal E}_n-{\cal E}_{n-1}}\,
  \phi_{n,n}(x;\lambda)
  =\sfrac{1}{n!}A_0^{\dagger}A_1^{\dagger}\cdots A_{n-1}^{\dagger}
  \phi_{0}(x;\lambda+\sfrac{n}{2}).
  \label{phinnn}
\end{equation}
This is another formula giving the eigenfunction
$\phi_n(x;\lambda)\propto P_n^{(\lambda)}(x)\phi_0(x;\lambda)$.
The situation is depicted in Fig.1. The operator $A$ acts to the right and
$A^\dagger$ to the left along the horizontal ({\em isospectral\/}) line.
They should not be confused with the  {\em annihilation\/}
and {\em creation\/} operators, which act along the vertical line of
a given Hamiltonian $H_s$ going from one energy level $n$ to another $n\pm1$.
The annihilation and creation operators will be discussed in section 
\ref{morex} in a more general setting.
\begin{figure}
 \centering
 \includegraphics*[scale=.7]{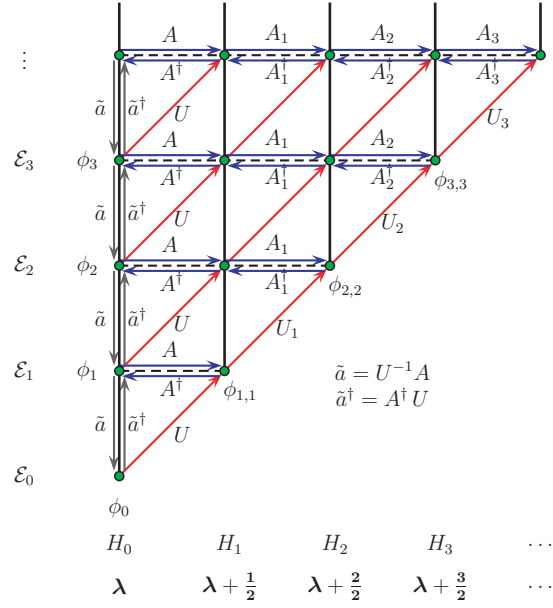}
 \caption{A schematic diagram of the energy levels and the associated
 Hamiltonian systems  together with the definition of the $A$ and
 $A^\dagger$ operators and the `creation'  and `annihilation' operators.
 The parameter set is indicated below each Hamiltonian.}
 \label{fig:aadagdef}
\end{figure}

\section{Other examples of shape invariant ``discrete quantum mechanics"}
\label{morex}
The other examples of {\em shape invariant\/} ``discrete quantum mechanics"
are related to a two-parameter deformation of the Hermite polynomial, and
two- and three-parameter deformation of the Laguerre polynomial. The
Laguerre polynomials are the eigenfunctions of a harmonic oscillator with a
centrifugal barrier ($1/x^2$ potential). They all belong to the Askey scheme
of hypergeometric orthogonal polynomials \cite{koeswart}. Demonstration of
the shape invariance and derivation of shift operators and eigenfunctions,
{\em etc.\/} go almost parallel with the
case of the deformed Hermite polynomials in section \ref{defher}.
We discuss them collectively by adopting an (almost) self-evident notation.

\begin{gather}
  \mbox{(\romannumeral1)}\,:\,
  \mbox{two-parameter deformation of the Hermite polynomial}, \nonumber\\
  \qquad\mbox{a special case of the {\bf continuous Hahn polynomial}},
  \nonumber\\
  \qquad V(x)=V(x;\bm{\lambda})=(a+ix)(b+ix),\\
  \qquad\phi_0(x)=\phi_0(x;\bm{\lambda})\propto |\Gamma(a+ix)\Gamma(b+ix)|,
  \label{phi01}\\
  \qquad\bm{\lambda}=(a,b),\quad 
  {\cal E}_n={\cal E}_n(\bm{\lambda})=\sfrac12n(n+2a+2b-1),
  \quad a,b\in\mathbb{R}_+,
  \label{eni}\\
  \qquad P_n(x)=P_n^{(\bm{\lambda})}(x)\propto p_n(x;a,b,a,b).\\
%
  \mbox{(\romannumeral2)}\,:\,
  \mbox{two-parameter deformation of the Laguerre polynomial}, \nonumber\\
  \qquad \mbox{the {\bf continuous dual Hahn polynomial}}, \nonumber\\
  \qquad V(x)=V(x;\bm{\lambda})=\frac{(a+ix)(b+ix)(c+ix)}{2ix(2ix+1)},
  \label{contdualhahn}\\
  \qquad\phi_0(x)=\phi_0(x;\bm{\lambda})
  \propto \left|\frac{\Gamma(a+ix)\Gamma(b+ix)\Gamma(c+ix)}
  {\Gamma(2ix)}\right|,\label{phi02}\\
%
  \qquad\bm{\lambda}=(a,b,c),\quad
  {\cal E}_n={\cal E}_n(\bm{\lambda})=\sfrac12n, \quad a,b,c\in\mathbb{R}_+,
  \label{enii}\\
  \qquad P_n(x)=P_n^{(\bm{\lambda})}(x)\propto S_n(x^2;a,b,c).
  \label{contdualhahnpoly}\\
%
  \mbox{(\romannumeral3)}\,:\,
  \mbox{three-parameter deformation of the Laguerre polynomial},  \nonumber\\
  \qquad \mbox{the {\bf Wilson polynomial}}, \nonumber\\
  \qquad V(x)=V(x;\bm{\lambda})=\frac{(a+ix)(b+ix)(c+ix)(d+ix)}{2ix(2ix+1)},
  \label{Wilsonpoly}\\
  \qquad
  \phi_0(x)=\phi_0(x;\bm{\lambda})
  \propto \left|\frac{\Gamma(a+ix)\Gamma(b+ix)\Gamma(c+ix)\Gamma(d+ix)}
  {\Gamma(2ix)}\right|,\label{phi03}\\
%
  \qquad\bm{\lambda}=(a,b,c,d),\;
  {\cal E}_n={\cal E}_n(\bm{\lambda})=\sfrac12n(n+a+b+c+d-1),\;
  a,b,c,d \in\mathbb{R}_+,
  \label{eniii}\\
  \qquad P_n(x)=P_n^{(\bm{\lambda})}(x)\propto W_n(x^2;a,b,c,d).
  \label{pn3}
\end{gather}
For space reasons we write the ground state
wavefunction in terms of the absolute value symbol as in (\ref{phi01}) which
should read
\begin{equation*}
  |\Gamma(a+ix)\Gamma(b+ix)|=
  \sqrt{\Gamma(a+ix)\Gamma(b+ix)\Gamma(a-ix)\Gamma(b-ix)}
\end{equation*}
as in (\ref{phi0form}).

\paragraph{Factorised Hamiltonian}
Factorisation and consequently the positive semi-definiteness
of the generic Hamiltonian
hold exactly the same as before:
\begin{gather}
  H\eqdef\sfrac12\sqrt{V(x)}\,e^{-i\partial_x}\sqrt{V^*(x)}
  +\sfrac12\sqrt{V^*(x)}\,e^{i\partial_x}\sqrt{V(x)}
  -\sfrac12(V(x)+V^*(x)),
  \label{geneH}\\
  H=A^{\dagger}A,
  \label{geneHfac}\\
  A=A(x;\bm{\lambda})
  \eqdef\frac{1}{\sqrt{2}}\Bigl(e^{-\frac{i}{2}\partial_x}\sqrt{V^*(x)}
  -e^{\frac{i}{2}\partial_x}\sqrt{V(x)}\Bigr),\\
  A^{\dagger}=A(x;\bm{\lambda})^{\dagger}
  \eqdef\frac{1}{\sqrt{2}}\Bigl(\sqrt{V(x)}\,e^{-\frac{i}{2}\partial_x}
  -\sqrt{V^*(x)}\,e^{\frac{i}{2}\partial_x}\Bigr).
\end{gather}
Here $\bm{\lambda}$ denotes the (set of) parameters.
The ground state $\phi_0$ is annihilated by $A$:
\begin{equation}
  A\phi_0(x)=0\quad
  \Longrightarrow H\phi_0(x)=0,\quad \mathcal{E}_0=0.
  \label{genphi0form}
\end{equation}
Verification of the ground state wavefunction (\ref{phi01}) for (i),
(\ref{phi02}) for (ii) and (\ref{phi03}) for (iii)
is straightforward.
The similarity transformed Hamiltonian
$\tilde{H}=\phi_0^{-1}\circ H\circ\phi_0$, (\ref{tilH}) determines the other
eigenfunctions of the Hamiltonian in the form
$\phi_n(x)\propto P_n(x)\phi_0(x)$, $\tilde{H}P_n=\mathcal{E}_nP_n$ 
(\ref{Htileq}).
They are the {\em difference equation\/} of the form
\begin{equation}
  V(x)P_n(x-i)+V^*(x)P_n(x+i)-(V(x)+V^*(x))P_n(x)=2\mathcal{E}_nP_n(x).
\end{equation}
Explicitly they read for the three cases listed above:
\begin{eqnarray}
  \hspace*{-8mm}\mbox{(\romannumeral1)}&:& (a+ix)(b+ix)P_n(x-i)+
 (a-ix)(b-ix)P_n(x+i)\nonumber\\
  \hspace*{-8mm}&&=2(ab-x^2+\mathcal{E}_n)P_n(x).\\
  \hspace*{-8mm}\mbox{(\romannumeral2)}&:&
  (a+ix)(b+ix)(c+ix)(2ix-1)P_n(x-i)+\nonumber\\
  \hspace*{-8mm}&&(a-ix)(b-ix)(c-ix)(2ix+1)P_n(x+i)
  \nonumber\\
  \hspace*{-8mm}&& =-2ix\left[(ab+ac+bc-2abc +
  2(a+b+c)x^2-x^2+2(4x^2+1)\mathcal{E}_n\right]P_n(x).\\
  \hspace*{-8mm}\mbox{(\romannumeral3)}&:&
  (a+ix)(b+ix)(c+ix)(d+ix)(2ix-1)P_n(x-i)+\nonumber\\
  \hspace*{-8mm}&& (a-ix)(b-ix)(c-ix)(d-ix)(2ix+1)P_n(x+i)
  \nonumber\\
  \hspace*{-8mm}&&=-2ix\left[(abc+abd+acd+bcd-2abcd +
  2(ab+ac+ad+bc+bd+cd)x^2\right.\nonumber\\
  \hspace*{-8mm}&&\qquad\qquad \left.
  -(a+b+c+d)x^2-2x^4+2(4x^2+1)\mathcal{E}_n\right]P_n(x).
\end{eqnarray}
They admit a degree $n$ polynomial solution in $x$ for (i),
and in $x^2$ for (ii) and (iii).
By comparing the coefficients of the leading degree, one obtains
the energy eigenvalues $\mathcal{E}_n$ (\ref{eni}), (\ref{enii}),
(\ref{eniii}).
The solutions form orthogonal polynomials satisfying three term
recurrence, which will not be shown here, see \cite{koeswart}.

The corresponding factorisation of $\tilde{H}$,
\begin{equation}
  \tilde{H}=BC,\quad
  C=\sfrac{i}{2}(e^{-\frac{i}{2}\partial_x}
  -e^{\frac{i}{2}\partial_x}),\quad
  B=-i(V(x)e^{-\frac{i}{2}\partial_x}
  -V^*(x)e^{\frac{i}{2}\partial_x}),
\end{equation}
provides the {\em forward\/} and {\em backward\/} shift operators
of the polynomials.
The $C$ operator, corresponding to the $A$ operator,
shifts to the right along the isospectral line:
$C: P_n^{(\bm{\lambda})}(x)\to
P_{n-1}^{(\bm{\lambda}+\bm{\sfrac{1}{2}})}(x)$ 
with $\bm{\sfrac12}=(\sfrac12,\sfrac12,\cdots)$.
The $B$ operator, corresponding to the $A^\dagger$ operator,
shifts to the left along the isospectral line:
$B: P_{n-1}^{(\bm{\lambda}+\bm{\sfrac{1}{2}})}(x)\to
P_{n}^{(\bm{\lambda})}(x)$.
For each case they are:
\begin{gather}
  \hspace*{-10mm}
  \mbox{(\romannumeral1)}\,:\,
  BCp_n(x;\bm{\lambda})=\mathcal{E}_np_n(x;\bm{\lambda})\nonumber\\
  \hspace*{-10mm}
  \qquad \Longleftarrow
  Cp_n(x;\bm{\lambda})=
  (\mathcal{E}_n/n)p_{n-1}(x;\bm{\lambda}+\bm{\sfrac{1}{2}}),
  \quad Bp_{n-1}(x;\bm{\lambda}+\bm{\sfrac{1}{2}})=np_n(x;\bm{\lambda}),\\
  \hspace*{-10mm}
  \qquad p_n(x+\sfrac{i}{2};\bm{\lambda})-p_n(x-\sfrac{i}{2};\bm{\lambda})=
  i(n+2a+2b-1)p_{n-1}(x;\bm{\lambda}+\bm{\sfrac{1}{2}}),\\
  \hspace*{-10mm}
  \qquad (a-ix)(b-ix)p_{n-1}(x+\sfrac{i}{2};\bm{\lambda}+\bm{\sfrac{1}{2}})
  -(a+ix)(b+ix)p_{n-1}(x-\sfrac{i}{2};\bm{\lambda}+\bm{\sfrac{1}{2}})
  \nonumber\\
  \hspace*{-10mm}
  \qquad\quad  =-inp_n(x;\bm{\lambda}).\\
%
  \hspace*{-10mm}
  \mbox{(\romannumeral2)}\,:\,
  BCS_n(x^2;\bm{\lambda})=\mathcal{E}_nS_n(x^2;\bm{\lambda})\nonumber\\
  \hspace*{-10mm}
  \qquad \Longleftarrow
  CS_n(x^2;\bm{\lambda})=
  -2\mathcal{E}_nxS_{n-1}(x^2;\bm{\lambda}+\bm{\sfrac{1}{2}}),\ 
  -2B\,xS_{n-1}(x^2;\bm{\lambda}+\bm{\sfrac{1}{2}})=S_n(x^2;\bm{\lambda}),\\
  \hspace*{-10mm}
  \qquad
  S_n((x+\sfrac{i}{2})^2;\bm{\lambda})-S_n((x-\sfrac{i}{2})^2;\bm{\lambda})=
  -2inxS_{n-1}(x^2;\bm{\lambda}+\bm{\sfrac{1}{2}}),\\
  \hspace*{-10mm}
  \qquad
  (a-ix)(b-ix)(c-ix)S_{n-1}((x+\sfrac{i}{2})^2;\bm{\lambda}+\bm{\sfrac{1}{2}})
  \nonumber\\
  \hspace*{-10mm}
  \qquad
  -(a+ix)(b+ix)(c+ix)S_{n-1}
  ((x+\sfrac{i}{2})^2;\bm{\lambda}+\bm{\sfrac{1}{2}})\nonumber\\
  \hspace*{-10mm}
  \qquad\quad  =-2ixS_n(x^2;\bm{\lambda}).\\
%
  \hspace*{-10mm}
  \mbox{(\romannumeral3)}\,:\,
  BCW_n(x^2;\bm{\lambda})=\mathcal{E}_nW_n(x^2;\bm{\lambda})\nonumber\\
  \hspace*{-10mm}
  \qquad \Longleftarrow
  CW_n(x^2;\bm{\lambda})=
  -2\mathcal{E}_nxW_{n-1}(x^2;\bm{\lambda}+\bm{\sfrac{1}{2}}),\ 
  -2B\,xW_{n-1}(x^2;\bm{\lambda}+\bm{\sfrac{1}{2}})=W_n(x^2;\bm{\lambda}),\\
  \hspace*{-10mm}
  \qquad
  W_n((x+\sfrac{i}{2})^2;\bm{\lambda})-
  W_n((x-\sfrac{i}{2})^2;\bm{\lambda})\nonumber\\
  \hspace*{-10mm}
  \qquad\quad =
  -2in(n+a+b+c+d-1)\,xW_{n-1}(x^2;\bm{\lambda}+\bm{\sfrac{1}{2}}),\\
  \hspace*{-10mm}
  \qquad
  (a-ix)(b-ix)(c-ix)(d-ix)
  W_{n-1}((x+\sfrac{i}{2})^2;\bm{\lambda}+\bm{\sfrac{1}{2}})
  \nonumber\\
  \hspace*{-10mm}
  \qquad
  -(a+ix)(b+ix)(c+ix)(d+ix)
  W_{n-1}((x+\sfrac{i}{2})^2;\bm{\lambda}+\bm{\sfrac{1}{2}})\nonumber\\
  \hspace*{-10mm}
  \qquad\quad  =-2ixW_n(x^2;\bm{\lambda}).
\end{gather}

\paragraph{Shape Invariance}
For shape invariance, we first need to find the operators $A_1$,
$A_1^\dagger$ and a real constant $\mathcal{E}_1$ satisfying
\begin{gather}
  H_1=AA^{\dagger}=A_1^{\dagger}A_1+{\cal E}_1,
  \label{genH1}\\
  A_1=A_1(x;\bm{\lambda})
  \eqdef\frac{1}{\sqrt{2}}\Bigl(e^{-\frac{i}{2}\partial_x}\sqrt{V_1^*(x)}
  -e^{\frac{i}{2}\partial_x}\sqrt{V_1(x)}\Bigr),\\
  A_1^{\dagger}=A_1(x;\bm{\lambda})^{\dagger}
  \eqdef\frac{1}{\sqrt{2}}\Bigl(\sqrt{V_1(x)}\,e^{-\frac{i}{2}\partial_x}
  -\sqrt{V_1^*(x)}\,e^{\frac{i}{2}\partial_x}\Bigr).
\end{gather}
In other words, given $V(x)=V(x;\bm{\lambda})$, find
a new potential $V_1(x)=V_1(x;\bm{\lambda})$ satisfying
\begin{gather}
  V_1^*(x-i)V_1(x)=V^*(x-\sfrac{i}{2})V(x-\sfrac{i}{2}),
  \label{V1eq1}\\
  V_1(x)+V_1^*(x)=V(x+\sfrac{i}{2})+V^*(x-\sfrac{i}{2})+2{\cal E}_1.
  \label{V1eq2}
  \end{gather}
If $V_1$ has the same form as $V$ with a shifted set of parameters
$\bm{\lambda}'$,
\begin{equation}
  V_1(x;\bm{\lambda})=V(x;\bm{\lambda}'),
\end{equation}
it is {\em shape invariant\/}.
Suppose $V_1$ has the form $V_1(x)=V(x-\sfrac{i}{2})g(x)$,
the above conditions
(\ref{V1eq1}), (\ref{V1eq2}) get slightly simplified:
\begin{gather}
  g^*(x-i)g(x)=1,
  \label{V1eq3}\\
  V(x-\sfrac{i}{2})g(x)+V^*(x+\sfrac{i}{2})g^*(x)
  =V(x+\sfrac{i}{2})+V^*(x-\sfrac{i}{2})+2{\cal E}_1.
  \label{V1eq4}
\end{gather}
The following choice satisfies the above conditions (\ref{V1eq3}),
(\ref{V1eq3}) for the three cases (\ref{phi01})--(\ref{pn3}):
\begin{eqnarray}
  \mbox{(\romannumeral1)}&:&
  g(x)=1,\quad \mathcal{E}_1(\bm{\lambda})=a+b,\\
  && V_1(x;\bm{\lambda})=V(x-\sfrac{i}{2};a,b)
  =V(x;a+\sfrac12,b+\sfrac12).\\
  \mbox{(\romannumeral2)}&:&
  g(x)=\frac{2ix+2}{2ix},\quad \mathcal{E}_1(\bm{\lambda})=\sfrac{1}{2},\\
  && V_1(x;\bm{\lambda})=V(x-\sfrac{i}{2};a,b,c)g(x)
  =V(x;a+\sfrac12,b+\sfrac12,c+\sfrac12).\\
  \mbox{(\romannumeral3)}&:&
  g(x)=\frac{2ix+2}{2ix},\quad 
  \mathcal{E}_1(\bm{\lambda})=\sfrac12(a+b+c+d),\\
  && V_1(x;\bm{\lambda})=V(x-\sfrac{i}{2};a,b,c,d)g(x)
  =V(x;a+\sfrac12,b+\sfrac12,c+\sfrac12,d+\sfrac12).
\end{eqnarray}
With $\bm{\sfrac12}=(\sfrac12,\sfrac12,\cdots)$, we have collectively
\begin{equation}
  \hspace*{-10mm}  
  V_1(x;\bm{\lambda})=V(x;\bm{\lambda}+\bm{\sfrac12}),\ 
  A_1(x;\bm{\lambda})=A(x;\bm{\lambda}+\bm{\sfrac12}),\ 
  H_1(x;\bm{\lambda})=
  H(x;\bm{\lambda}+\bm{\sfrac12})+{\cal E}_1(\bm{\lambda}).
\end{equation}

Starting from $V_0=V$, $H_0=H$, $\phi_{0,n}=\phi_n$, let us define
$V_s$, $H_s$, $\phi_{s,n}$ ($n\geq s\geq 0$) step by step:
\begin{gather}
  V_{s+1}(x;\bm{\lambda})\eqdef V_s(x;\bm{\lambda}+\bm{\sfrac12}),\\
  H_{s+1}(x;\bm{\lambda})\eqdef
  A_s(x;\bm{\lambda})A_s(x;\bm{\lambda})^{\dagger}
  +{\cal E}_s(\bm{\lambda}),\\ 
  \phi_{s+1,n}(x;\bm{\lambda})\eqdef
  A_s(x;\bm{\lambda})\phi_{s,n}(x;\bm{\lambda}),\\
{\cal E}_{s}(\bm{\lambda})={\cal E}_{s-1}(\bm{\lambda})
+{\cal E}_1(\bm{\lambda}+\bm{(s-1)/{2}}).
\end{gather}
Here
$A_s$, $A_s^{\dagger}$ are defined by
\begin{gather}
  A_s(x;\bm{\lambda})
  \eqdef\frac{1}{\sqrt{2}}\Bigl(
  e^{-\frac{i}{2}\partial_x}\sqrt{V_s^*(x;\bm{\lambda})}
  -e^{\frac{i}{2}\partial_x}\sqrt{V_s(x;\bm{\lambda})}\Bigr),\\
  A_s(x;\bm{\lambda})^{\dagger}
  \eqdef\frac{1}{\sqrt{2}}\Bigl(
  \sqrt{V_s(x;\bm{\lambda})}\,e^{-\frac{i}{2}\partial_x}
  -\sqrt{V_s^*(x;\bm{\lambda})}\,e^{\frac{i}{2}\partial_x}\Bigr).
\end{gather}
As a consequence of the shape invariance, we obtain for $n\ge s\ge0$:
\begin{gather}
  V_s(x;\bm{\lambda})=V(x;\bm{\lambda}+\bm{\sfrac{s}{2}}),\\
  A_s(x;\bm{\lambda})=A(x;\bm{\lambda}+\bm{\sfrac{s}{2}}),\quad
  A_s(x;\bm{\lambda})^{\dagger}
  =A(x;\bm{\lambda}+\bm{\sfrac{s}{2}})^{\dagger},\\
  H_s(x;\bm{\lambda})
  =A_s(x;\bm{\lambda})^{\dagger}A_s(x;\bm{\lambda})
  +{\cal E}_s(\bm{\lambda}),\\
  H_s(x;\bm{\lambda})\phi_{s,n}(x;\bm{\lambda})
  ={\cal E}_n(\bm{\lambda})\phi_{s,n}(x;\bm{\lambda}),\\
  A_s(x;\bm{\lambda})\phi_{s,s}(x;\bm{\lambda})=0,\\
  A_s(x;\bm{\lambda})^{\dagger}\phi_{s+1,n}(x;\bm{\lambda})
  =({\cal E}_n(\bm{\lambda})-{\cal E}_s(\bm{\lambda}))
  \phi_{s,n}(x;\bm{\lambda}).
  \end{gather}
The eigenfunction $\phi_{s,n}$ has $n-s$ nodes,
or the corresponding polynomial is of degree $n-s$ 
in $x$ for (i), and in $x^2$ for (ii) and (iii).
In the latter case, the zeros on a half line, 
say $x\ge0$, count as the nodes.

{}From these we obtain formulas
\begin{gather}
  \phi_{s,n}(x;\bm{\lambda})=A_{s-1}(x;\bm{\lambda})\cdots
  A_1(x;\bm{\lambda})A_0(x;\bm{\lambda})\phi_n(x;\bm{\lambda}),
  \label{genphinsform}\\
  \phi_n(x;\bm{\lambda})=
  \frac{A_0(x;\bm{\lambda})^{\dagger}}
  {{\cal E}_n(\bm{\lambda})-{\cal E}_0(\bm{\lambda})}\,
  \frac{A_1(x;\bm{\lambda})^{\dagger}}
  {{\cal E}_n(\bm{\lambda})-{\cal E}_1(\bm{\lambda})}\cdots
  \frac{A_{n-1}(x;\bm{\lambda})^{\dagger}}
  {{\cal E}_n(\bm{\lambda})-{\cal E}_{n-1}(\bm{\lambda})}\,
  \phi_{n,n}(x;\bm{\lambda}),
   \label{genphinform}
  \end{gather}
corresponding to (\ref{phinsform}) and (\ref{phinnn})
discussed in section \ref{defher}.
The former (\ref {genphinsform}) gives the eigenfunction $\phi_{s,n}$ of the
$s$-th Hamiltonian $H_s$ along
the isospectral line with energy $\mathcal{E}_n$,
starting from $\phi_n$ of the original Hamiltonian $H$ by repeated application
of the $A$ operators. The latter (\ref {genphinform}), on the other hand,
expresses the $n$-th eigenfunction $\phi_n$ of the original Hamiltonian,
starting from the explicitly known ground state $\phi_{n,n}$ of the $n$-th
Hamiltonian $H_n$ by repeated application of the $A^\dagger$ operators.
The latter formula (\ref{genphinform}), a simple generalisation of the
well-known formula for the harmonic oscillator $|n\rangle\propto
(a^\dagger)^n|0\rangle$, could also be understood as the generic form of the
Rodrigue's formula for the orthogonal polynomials.

In order to define  the annihilation and creation operators, let us introduce
normalised basis $\{\hat{\phi}_{s,n}\}_{n\ge s}$ for each Hamiltonian
$H_s$. Ordinarily, the phase of each element of an orthonormal basis could
be  completely arbitrary. In the present case, however, the eigenfunctions
are orthogonal polynomials. That is, they are real and the relations among
different degree members are governed by the three term recurrence
relations. So the phases of $\{\hat{\phi}_{s,n}\}_{n\ge s}$ are fixed.
Let us introduce a unitary (in fact  an orthogonal) operator $U_s$ mapping
the $s$-th orthonormal basis $\{\hat{\phi}_{s,n}\}_{n\ge s}$ to the
$(s+1)$-th 
$\{\hat{\phi}_{s+1,n}\}_{n\ge s+1}$ (see Fig. 1 and for example 
\cite{spivinzhed,kindao}):
\begin{equation}
  U_s\hat{\phi}_{s,n}=\hat{\phi}_{s+1,n+1},\quad
  U_s^\dagger \hat{\phi}_{s+1,n+1}=\hat{\phi}_{s,n}.
\end{equation}
We denote that $U_0=U$.
Roughly speaking $U$ increases the parameters from $\bm{\lambda}$ to
$\bm{\lambda}+\bm{\sfrac12}$:
\begin{equation}
  U: P_n^{(\bm{\lambda})}(x)\to  P_n^{(\bm{\lambda}+\bm{\sfrac12})}(x),
  \quad
  U^\dagger: P_n^{(\bm{\lambda}+\bm{\sfrac12})}(x)\to
  P_n^{(\bm{\lambda})}(x).\nonumber
\end{equation}
Let us introduce an annihilation $\tilde{a}$ and a creation
operator
$\tilde{a}^\dagger$  for the Hamiltonian $H$ as follows:
\begin{equation}
  \tilde{a}=\tilde{a}(x;\bm{\lambda})\eqdef
  U^{\dagger}A(x;\bm{\lambda}),
  \quad
  \tilde{a}^{\dagger}=\tilde{a}(x;\bm{\lambda})^{\dagger}\eqdef
  A(x;\bm{\lambda})^{\dagger}U.
\end{equation}
It is straightforward to derive
\begin{eqnarray}
  &&H=\tilde{a}^{\dagger}\tilde{a},\\
  &&[\tilde{a},\tilde{a}^{\dagger}]\hat{\phi}_n(x;\lambda)
  =({\cal E}_{n+1}(\bm{\lambda})-{\cal E}_n(\bm{\lambda}))
  \hat{\phi}_n(x;\lambda).
\end{eqnarray}
For the harmonic oscillator (\ref{osciham}) $U=$id. and we
recover the known result.
For the linear spectrum $\mathcal{E}_n=n,n/2$ of the
deformed Hermite polynomial (\ref{H}) and the two parameter deformation
of the Laguerre polynomial (\ref{contdualhahn})--(\ref{contdualhahnpoly}),
$\tilde{a}$ and
$\tilde{a}^\dagger$ have the same (up to rescaling)  commutation relations as
those of the harmonic oscillator.
Essentially the same arguments and
results hold for the annihilation $\tilde{a}_s=U_s^\dagger
A_s(x;\bm{\lambda})$ and the creation  operator
$\tilde{a}_s^\dagger=A_s(x;\bm{\lambda})U_s$  for the Hamiltonian
$H_s$.

\section{Dynamical Background}
Here we will show briefly a logical (dynamical) path
that led to the shape invariant difference equations introduced
in section \ref{defher}.
The equilibrium position of an $n$-particle A-type Calogero
system is determined by
\begin{equation}
  \sum_{k=1\atop k\ne j}^n\frac{1}{x_j-x_k}
  =x_j\,,\quad j=1,\ldots,n,
\label{stirel}
\end{equation}
after adjustment of the coupling constants and rescaling of the
variables \cite{calmat,sti}. They describe the zeros of the Hermite
polynomial, since $H_n(x)\eqdef  2^n\prod_{j=1}^n(x-x_j)$ satisfies the
differential equation
\begin{equation}
  H_n^{''}(x)-2xH_n'(x)+2nH_n(x)=0
  \label{hereq}
\end{equation}
and the three term recurrence \cite{sti} $n\ge0$,
\begin{equation}
  H_{n+1}(x)-2xH_n(x)+2nH_{n-1}(x)=0,\quad H_0=1,\quad H_{-1}=0.
  \label{herthr}
\end{equation}
The equilibrium position of an $n$-particle A-type van Diejen
system (or the rational R-S system with a linear confining potential)
is determined by
\begin{equation}
  \prod_{k=1\atop k\ne j}^n
  \frac{x_j-x_k-i\sqrt{\delta}}{x_j-x_k+i\sqrt{\delta}}=
  \frac{1-i\sqrt{\delta}\,x_j}{1+i\sqrt{\delta}\,x_j}\,,
  \quad j=1,\ldots,n,
  \label{rclAeq2}
\end{equation}
after adjustment of the coupling constants and rescaling of the
variables \cite{rsos2}. The corresponding {\em deformed\/} polynomial
$H_n(x,\delta)\eqdef  2^n\prod_{j=1}^n(x-x_j)$ satisfies the
difference equation
\begin{equation}
  \hspace*{-4mm}
  \Bigl(x+\frac{i}{\sqrt{\delta}}\Bigr)H_n(x+i\sqrt{\delta},\delta)
  -\Bigl(x-\frac{i}{\sqrt{\delta}}\Bigr)H_n(x-i\sqrt{\delta},\delta)
  =\frac{2i}{\sqrt{\delta}}(1+n\delta\,)H_n(x,\delta),
  \label{feq_rclA}
\end{equation}
and the three term recurrence \cite{rsos2}  $n\ge0$,
\begin{equation}
  \hspace*{-4mm}H_{n+1}(x,\delta)-2xH_n(x,\delta)
  +\Bigl(2n+n(n-1)\delta\Bigr)H_{n-1}(x,\delta)=0,\quad H_0=1,\quad
  H_{-1}=0.
  \label{rec_rclA2}
\end{equation}
It is elementary to show that (\ref{rclAeq2}), (\ref{feq_rclA}),
(\ref{rec_rclA2}) reduce to (\ref{stirel}), (\ref{hereq}), (\ref{herthr})
in the zero deformation limit $\delta\to0$; $\lim_{\delta\to0}H_n(x,\delta)=H_n(x)$.
The relationship between the deformed Hermite polynomial and the
special case of the Meixner-Pollaczek polynomial discussed in section
\ref{defher} is
\begin{equation}
  H_n(x,\delta)=n!\sqrt{\delta}^{\,n}P_n^{(\frac{1}{\delta})}
  \Bigl(\frac{x}{\sqrt{\delta}}\,;\frac{\pi}{2}\Bigr).
\end{equation}
It is instructive to write down the classical Hamiltonian of the
$n$-particle A-type van Diejen system
\begin{gather}
  H(p,q)=\sum_{j=1}^n\left( \cosh p_j\,\sqrt{V_j(q)\,{V}_j^*(q)}
  -\frac12\Bigl(V_j(q)+{V}_j^*(q)\Bigr)\right),\\
  V_j(q)=\left(1+\frac{iq_j}{a}\right)\prod_{k=1\atop k\ne j}^n
  \left(1-\frac{ig}{q_j-q_k}\right),
  \label{Ham}
\end{gather}
in which $a$ and $g$ are coupling constants.
In fact, the Hamiltonian (\ref{H}) is the single particle ($n=1$)
case with $p_j=-i{\partial}/{\partial q_j}$ and proper ordering.
We refer to \cite{rsos2} for similar orientation of the dynamical
systems introduced in section \ref{morex}.

\section{Comments and Discussion}
Several examples of shape invariant difference equations are discussed
in some detail.
They are related to deformation of the harmonic oscillator
without/with a centrifugal potential and their eigenfunctions are
deformed Hermite and Laguerre polynomials belonging to
the Askey-scheme of hypergeometric orthogonal polynomials.
They arise in the problems of determining the equilibrium positions
of the Ruijsenaars-Schneider-van Diejen systems, which are integrable
deformation of the celebrated Calogero-Sutherland systems of
exactly solvable multi-particle quantum mechanics.
Here we have treated rational potentials $V(x)$ only.
Obviously the method works for a wider range of potentials,
the trigonometric, hyperbolic, elliptic, {\em etc\/}.
We have not discussed the shape invariance of the discrete
systems corresponding to the equilibria of the trigonometric
Ruijsenaars-Schneider systems.
These  have various kinds of deformed Jacobi polynomials as
eigenfunctions \cite{rsos2}.
We will report on this subject elsewhere.

In the literature, shape invariance is discussed almost always within
the context of `supersymmetric quantum mechanics'\cite{genden}.
We presume that this might be  the psychological barrier
for considering the shape invariance in discrete quantum mechanics,
or shape invariant difference equations.
As is well-known, supersymmetry is the `square root' of
the space-time symmetry, say the Poincar\'e symmetry, which is
usually lost if the space-time is discretised.....
In fact, it is important to realise that main results of
`supersymmetric quantum mechanics' are already contained in Crum's theorem
without supersymmetry or shape invariance.
Among them are: the existence of {\em associated isospectral\/}
Hamiltonians $H_1$, \ldots, $H_s$, \ldots, as many as the number of discrete
levels of $H_0$ and the formulas of their eigenfunctions. For example,
$\phi_{s,n}$ of $H_s$ with the eigenvalue $\mathcal{E}_n$
is expressed in terms of the eigenfunctions
$\phi_0$, $\phi_1$,\ldots, $\phi_{s-1}$, $\phi_n$ of $H_0$ (Fig. 1):
\begin{gather}
  \phi_{s,n}=\frac{\mathcal{W}_{s,n}}{\mathcal{W}_s}, \quad n\ge s\ge1,\\
  \mathcal{W}_s=\mathcal{W}[\phi_0,\ldots,\phi_{s-1}],\quad
  \mathcal{W}_{s,n}=\mathcal{W}[\phi_0,\ldots,\phi_{s-1},\phi_n].
\end{gather}
Here the Wronskian $\mathcal{W}$  is defined as usual
$\mathcal{W}[f_1,\ldots,f_n]=\det\left(f_j^{(i-1)}\right)_{1\le i,j\le n}$.
This formula, translated in the discrete dynamics context, corresponds
to (\ref{genphinsform}).

We do believe dynamical (Hamiltonian) interpretation of discrete
quantum systems (difference equations) would be useful and fruitful.
One of our starting points, equations determining a certain equilibrium,
{\em eg.\/} (\ref{rclAeq2}), are called Bethe ansatz like equations.
The present problems could be considered as those of finding associated 
polynomial solutions of Bethe ansatz like equations.
They  occur in many branches of theoretical
physics, for example, the quasi-exactly  solvable single and multi-particle
quantum systems \cite{muwst1} on top of the well-known integrable spin
chains \cite{ismail}.

\section*{Acknowledgements}
We are grateful to Norbert Euler, the Editor of JNMP, for the kind invitation
to contribute for this special issue in honour of Francesco Calogero.
We are very thankful to Francesco for many splendid gifts he presented to the
theoretical/mathematical physics community.
R.\,S. recalls with warmth and gratitude many nice occasions of academic
interactions with Francesco for over twenty years which have never failed
to be accompanied by his personal charm.
S. O. and R. S. are supported in part by Grant-in-Aid for Scientific
Research from the Ministry of Education, Culture, Sports, Science and
Technology, No.13135205 and No. 14540259, respectively.


\label{lastpage}

\end{document}